\documentclass[]{llncs}
\usepackage{graphicx}
\usepackage{subfigure}
\usepackage{fancyvrb}
\newtheorem{fact}{Fact}

\begin{document}



\title{Algorithms for Visualizing Phylogenetic Networks}
\author{Ioannis G. Tollis \inst{1} and Konstantinos  G. Kakoulis \inst{2}}
\institute{
	Department of Computer Science,  
	University of Crete, Greece.
	\and
	Department of Mechanical and Industrial Design Engineering, \\
	T.E.I. of West Macedonia, Greece.
	}
\maketitle


\begin{abstract}
We study the problem of visualizing phylogenetic networks, which are extensions
of the \textit{Tree of Life} in biology.
We use a space filling visualization method, called DAGmaps, in order to obtain
clear visualizations using  limited space.
In this paper, we restrict our attention to galled trees and galled networks and present linear time
algorithms for visualizing them as DAGmaps. 
\end{abstract}

\section{Introduction}

\footnotetext{Emails: tollis@csd.uoc.gr and kkakoulis@teiwm.gr }
The quest of the \textit{Tree of Life} arose centuries ago, and one of the first illustrations of an evolutionary tree was produced by Charles Darwin in 1859, in his book ``\textit{The Origin of Species}''. Over a century later, evolutionary biologists still used \textit{phylogenetic trees} to depict evolution.
A \textit{phylogenetic tree $T$ on $X$} is obtained by labeling the leaves of a tree by the set of taxa $X = \{x_1, x_2, \ldots, x_n\}$. Each taxon $x_i$ represents a species or an organism. 


The branches of the phylogenetic trees represent the evolution of species, and sometimes the length of their edges is scaled in order to represent the time.


As pointed out in \cite{Doolittle99}, molecular phylogeneticists were failing to find the true tree of life, not because their methods were inadequate or because they had chosen the wrong genes, but perhaps because the history of life cannot be properly represented as a tree. Indeed, the mechanisms of horizontal gene transfer, hybridization and genetic recombination necessitate the use of \textit{phylogenetic} network models to illustrate them.

There are many different types of phylogenetic networks which can be separated in two main classes according to \cite{HusonK07}: \textit{implicit} phylogenetic networks that provide tools to visualize and analyze incompatible phylogenetic signals, such as split networks \cite{Huson06}, and \textit{explicit} phylogenetic networks that provide explicit scenarios of reticulate evolution, such as hybridization networks \cite{Linder04,maddison1997gene}, horizontal gene transfer networks \cite{Hallett04} and recombination networks \cite{GusfieldEL03,Huson11}.\\


Visualization of phylogenetic trees and networks is an important part of this area, 
since most of these graphs are huge.
Furthermore, the usual node-link representation leads to visual clutter. 
Thus, alternative visualization of phylogenetic trees, such as treemaps, may be preferable.

Treemaps \cite{Johnson91}, a space filling technique for visualizing large hierarchical data sets,   display trees as a set of nested rectangles. 
The (root of the) tree is the initial rectangle.  Each subtree  is assigned to a
subrectangle, which is then tiled into smaller rectangles representing further subtrees.
Space filling visualizations, such as treemaps, have the capacity to display 
thousands of items legibly in limited space via a two dimensional map. 
Treemaps have been used in bioinformatics to visualize phylogenetic trees  \cite{Arvelakis05}, 
gene expression data \cite{mcconnell2002applications},  
gene ontologies \cite{Baehrecke2004,Symeonidis2006,Tao2004237}, 
and the Encyclopedia of Life [1].
An extension of treemaps is presented in \cite{Tsiaras09}, which manages to visualize not only trees, but also Directed Acyclic Graphs (DAGs). As shown in \cite{Tsiaras09}, it is not always possible to visualize a DAG with a DAGmap without having node duplications. 


In this paper we present space filling techniques that use DAGmap drawings for the visualization of two categories of phylogenetic networks, galled trees and planar galled networks. No node duplications appear in both visualization algorithms that we present.
%
%
In Section \ref{sectionDAGmap4GTN} 
we introduce an algorithm which locates the galls of a graph and examines whether this graph is a galled tree or a galled network. 
In Section \ref{dagmaps4GT} we describe how to draw the DAGmaps of galled trees, and we examine 
whether the galled trees and galled networks can be one-dimensionally DAGmap drawn. Finally, in Section \ref{dagmaps4GN} we 
present an algorithm for producing DAGmap drawings of planar galled networks.


\section{Preliminaries}
\label{sectionDAGmap4GTN}



Let $G = (V,E)$ be a directed graph (digraph) with $n = |V|$ nodes and
$m =  |V|$ edges. 
If $e = (u, v) \in E$ is a directed edge, we say
that $e$ is incident from $u$ (or outgoing from $u$) and incident to $v$ (or incoming
to $v$); edge $u$ is the origin of $e$ and node $v$ is the destination of $e$. 
A directed acyclic graph (DAG) is a digraph that contains no cycles.
A source of digraph $G$ is a node without incoming edges. 
A sink of $G$ is a node without outgoing edges. An internal node
of $G$ has both incoming and outgoing edges.

A drawing of a graph $G$ maps each node
$v$ to a distinct point of the plane and each edge $(u, v)$ to a simple open Jordan
curve, with endpoints $u$ and $v$. A drawing is planar if no two edges intersect
except, possibly, at common endpoints. A graph is planar if it admits a planar
drawing. Two planar drawings of a graph are equivalent if they determine the
same circular ordering of the edges around each node. An equivalence class of
planar drawings is a (combinatorial) embedding of $G$. An embedded graph is a
graph with a specified embedding. A planar drawing partitions the plane into
topologically connected regions that are called faces.

An upward drawing of a digraph is such that all the edges are represented by
directed curves increasing monotonically in the vertical direction. A digraph has an
upward drawing if and only if it is acyclic. A digraph is upward planar if it admits
a planar upward drawing. Note that a planar acyclic digraph does not necessarily
have a planar upward drawing. 
%
A graph is layered planar if it can be drawn such that the nodes are placed in horizontal rows or layers, the edges are drawn as polygonal chains connecting their end nodes, and there are no edge crossings.

In a phylogenetic network there can be three kind of nodes: \textit{root}, \textit{tree}, and \textit{reticulation} nodes.
A root node has no incoming edges. There is only one root node in every rooted phylogenetic network.
Tree nodes have exactly one ancestor.
Reticulation nodes have more than one ancestors.
It is easy to realize that a phylogenetic tree is a phylogenetic network without reticulation nodes.

In addition, there can be two kind of edges: \textit{tree}, and \textit{reticulation} edges.
A tree edge leads to a node that has exactly one incoming edge.
A reticulation edge leads to a node that has more than one incoming edges.

\textit{Reticulation} cycles are defined as follows. Since there is only one root node in every rooted phylogenetic network, in the corresponding undirected graph every reticulation node belongs to a cycle. This cycle, in the directed graph, is called reticulation cycle.

\begin{figure} []
	\centering
	\includegraphics[scale=0.4,clip = true, viewport = 0 0 400 490]{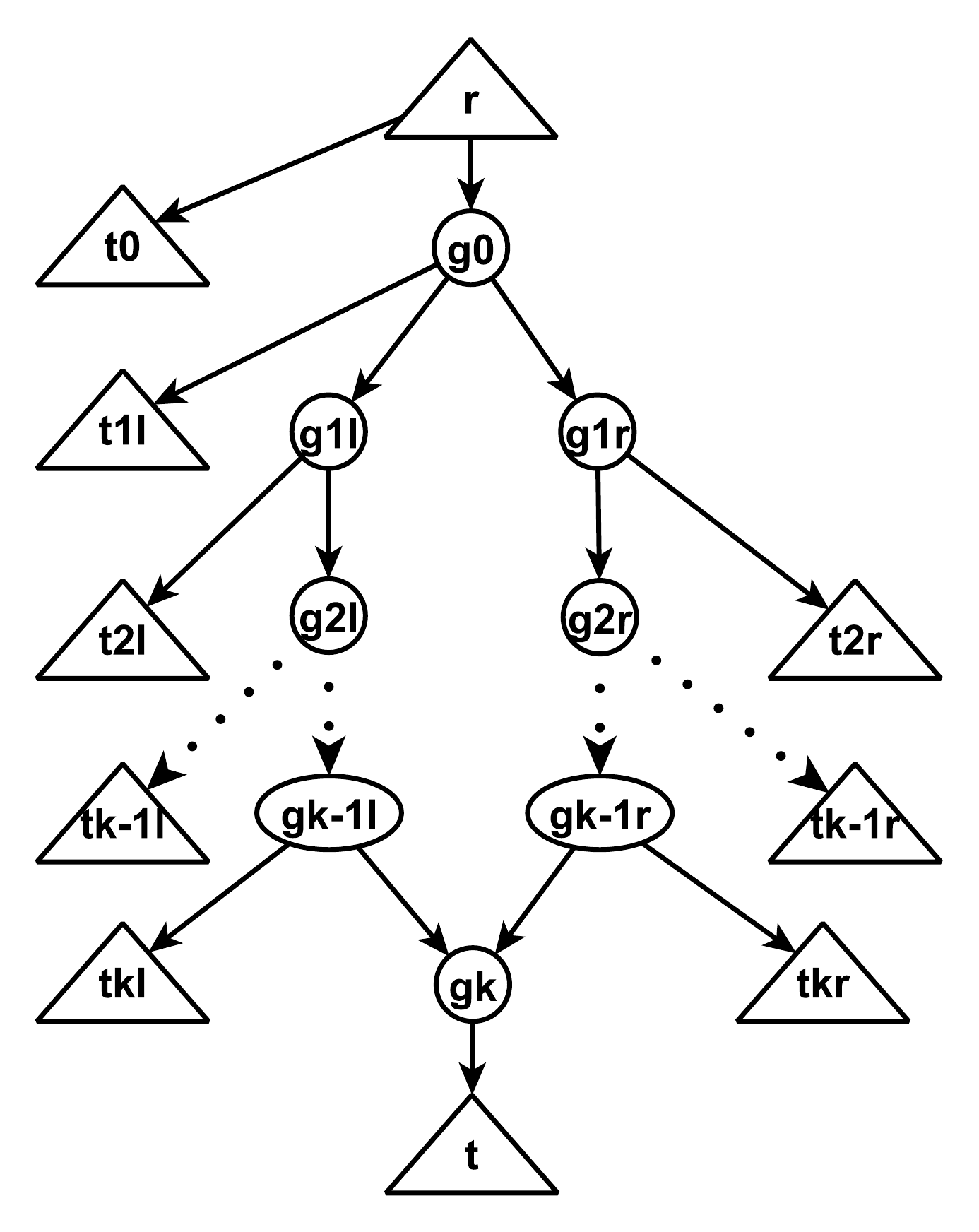}
	\caption{The structure of a gall. }
	\label{galled_tree}
\end{figure}

A \textit{gall} is a reticulation cycle in a phylogenetic network that shares no nodes with any other reticulation cycle.
It consists of a beginning node $g_0$, two chains (the left and the right one) and a reticulation node $g_k$, as shown in Figure \ref{galled_tree}.
The beginning node $g_0$ is on level $1$ of this subgraph, the reticulation node on level $k+1$, and the chain nodes are on the $i$ levels, $i\in\{2, \ldots, k\}$. Every level $i$ may contain either one or two chain nodes.
Every node $g_i$, $i\in\{0, \ldots, k\}$, of the gall may have a subtree $t_{i+1}$ as a descendant. These subtrees do not have more connections with this gall, because in that case a reticulation cycle would be created, which would share a node with the gall, and this is not allowed according to the definition of a gall.

A \textit{galled} tree is a phylogenetic network whose reticulation cycles are galls \cite{GusfieldEL03,Wang01}.
This is called the \textit{galled  tree  condition}.
Considering the definition of a gall, it is easy to realize that the reticulation nodes of a galled tree have indegree two. 

A \textit{Galled} network is a rooted phylogenetic network in which every reticulation cycle shares no reticulation nodes with any other reticulation cycle \cite{HusonRBGP09}.
This is called the \textit{galled  network  condition}.

In contrast to the galled trees, galled networks allow the reticulation cycles to share nodes, as long as they are not reticulation nodes. These reticulation cycles are called \textit{loose galls}. In the rest of the paper, whenever we refer to loose galls of a galled network, we will use the term \textit{galls} for simplification.

Galled trees \cite{GusfieldEL03,Jansson201466,Nakhleh2004,Wang01} and galled networks \cite{HusonK07,HusonRBGP09,Huson-S-11,Jansson2006} 
have received much attention in recent years.
They are important types of phylogenetic networks due to their biological significance and 
their simple, almost treelike, structure. 
A galled tree or network may suffice to accurately describe an evolutionary process 
when the number of recombination events is limited and most of them have occurred recently \cite{GusfieldEL03}.


\begin{figure}[]
	\centering
	\subfigure[]{
		\label{DAGmap1}
		\includegraphics[scale=0.25,clip = true, viewport = 150 100 650 500]{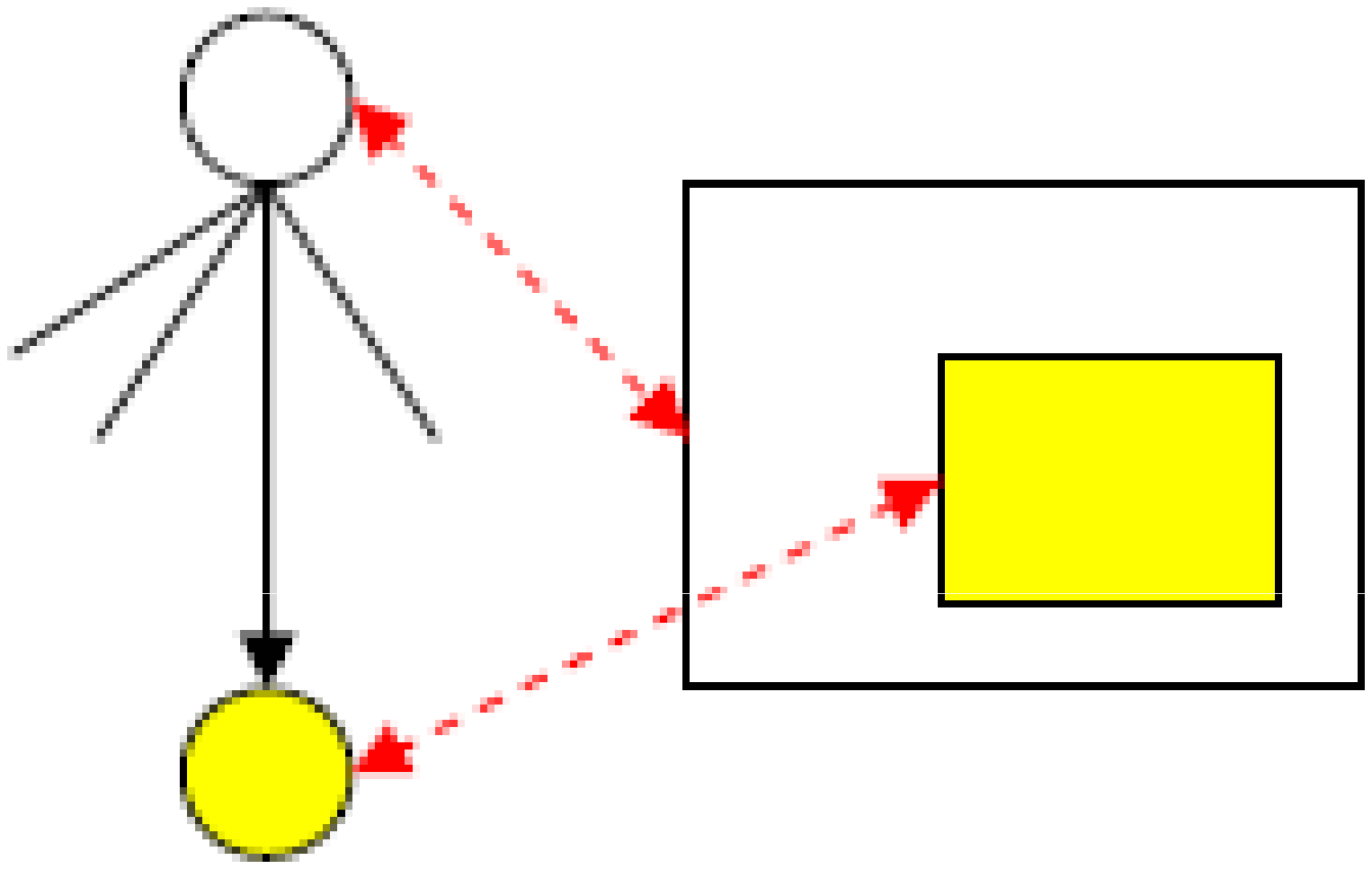}}
	\subfigure[]{
		\label{DAGmap2}
		\includegraphics[scale=0.25,clip = true, viewport = 100 100 750 520]{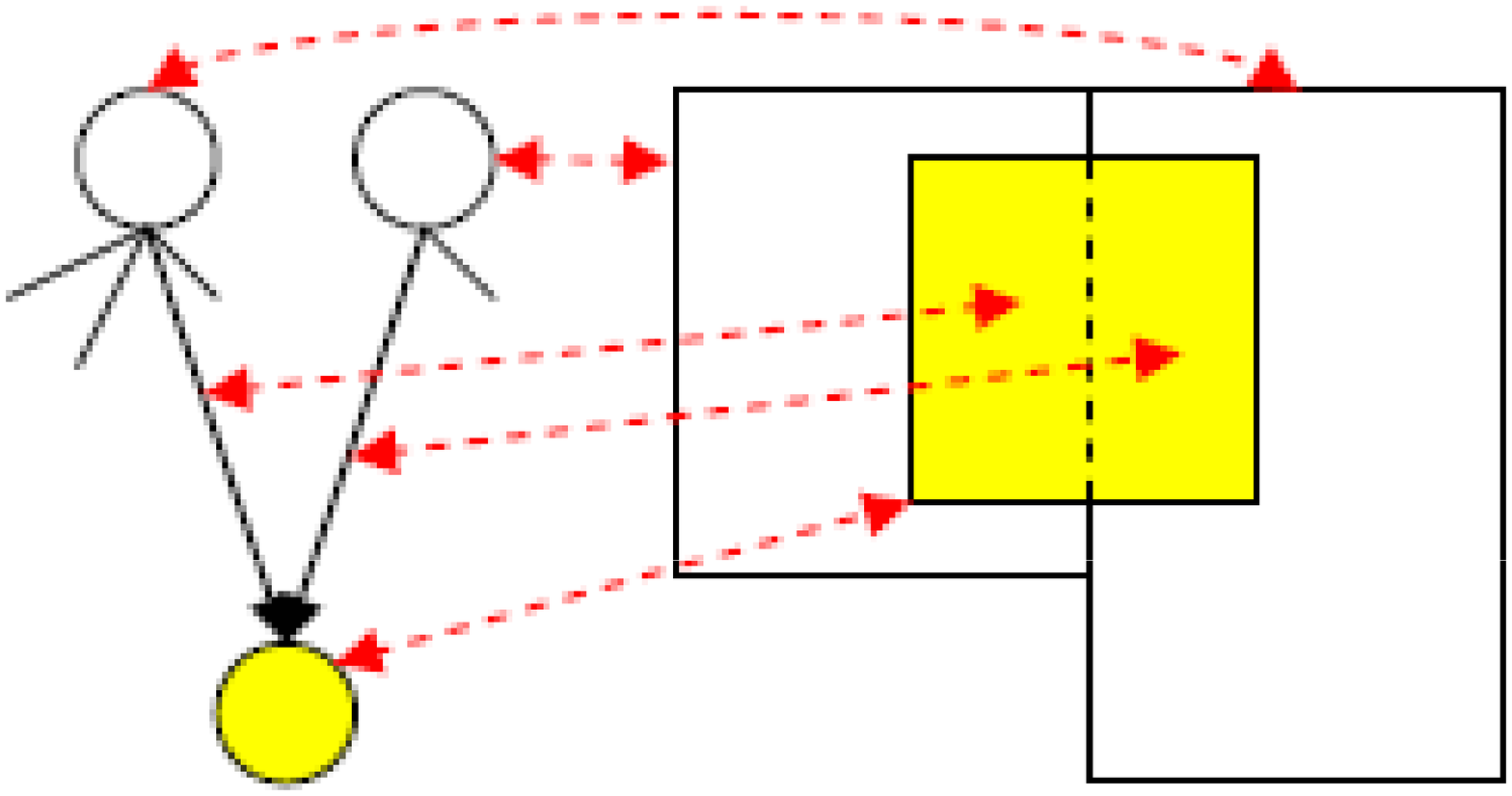}}
	\caption{({\bf a}) A treemap drawing. ({\bf b}) A DAGmap Drawing. }
	\label{DAGmap12}
\end{figure} 


\subsection{The DAGmap Problem}
\label{TheDAGmapProblem}

DAGmaps are space filling visualizations of DAGs that generalize
treemaps \cite{Tsiaras09}. 
%
The main properties of DAGmaps are shown in Figure \ref{DAGmap12}. 
In treemaps the rectangle of a child node is included into the rectangle 
of its parent node (see Figure \ref{DAGmap1}). In DAGmaps the rectangle of a node is included into 
the union of rectangles of its ancestors. 
Also the rectangle of an edge is contained in the intersection of the 
rectangles of its source and destination nodes (see Figure \ref{DAGmap2}).

The DAGmap problem is the problem of deciding whether a graph admits a DAGmap drawing without node duplications.
 Deciding whether or not a DAG admits a DAGmap drawing is NP-
 complete \cite{Tsiaras09}.
 Furthermore, the DAGmap problem remains NP-complete even when the graphs are  restricted to be galled networks: 
   

\begin{theorem}
	The DAGmap problem for galled networks is NP-complete.
	\label{NP}
\end{theorem}
{\bf Proof:} Omitted due to space limitations.  \hspace*{\fill} \framebox


\subsection{Locating the Galls}
\label{locateGalls}

%

The first task is to recognize whether a given phylogenetic network is a galled tree or a galled network.
Since they both contain galls, we will need to locate the galls
of the given phylogenetic network. This will allow us to check whether our network is a galled tree, a galled network, or none of them.  This can be accomplished by the following algorithm:

\noindent \textbf{Algorithm 1: Locating the galls of a graph}\\
\textbf{Input: } A Graph $G$.\\
\textbf{Output: } The set of galls and the characterization of $G$ as a galled tree or a galled network, or null if the graph is neither of them.
\begin{Verbatim}[]
1. Perform a simple graph traversal in order to locate the
   reticulation nodes.
2. If a node with more than two incoming edges is found, then
   return null.
3. For every reticulation node find its two parents. Each of these
   parents belongs to a chain of the gall.
4.   For every parent find its parent and assign it to the same
     chain. (At each step discover one node from each chain.)
5.   Continue this process until a node is found which already
     belongs to the other chain. This is the beginning node of the
     gall. If no such node is found, return null.
6. After locating all the galls, test the galled tree and the
   galled network condition.
7. If the galled tree condition holds then characterize the graph
   as a "galled tree".
8. Else if the galled network condition holds then characterize
   the graph as a "galled network".
9. Else return null.
10.Return the located galls.
\end{Verbatim}





This process will discover all the galls of the graph, since every reticulation node corresponds to exactly one gall. In addition, every chain node will be visited a constant number of times if we use a hash table to store the chain nodes. Also, the property that every gall has exactly one reticulation node guarantees that this algorithm will neither leave any gall undiscovered, nor claim to discover a gall that does not exist. Thus, it is straightforward to show that Algorithm 1 runs in $O(n+m)$ time.


\section{DAGmaps for Galled Trees}
\label{dagmaps4GT}

In this section we present techniques for drawing galled trees as DAGmaps.

\subsection{Drawing Galled Trees as DAGmaps}

Next, we present a three step algorithm for drawing galled trees as DAGmaps. 
First, we transform the input galled tree into a tree by collapsing the two chains of each gall into a single chain.
Then, we use treemap techniques to draw the tree. 
Finally, we expand the collapsed galls.
Next, we make some interesting observations:
\begin{fact}
	Any node of a galled tree has indegree at most two.
\end{fact}


If there were a node with indegree more than two in a galled tree, then this node would belong to more than one reticulation cycles, which means that there would be (more than one) reticulation cycles.

\begin{fact}
	Every galled tree is planar.
\end{fact}

\begin{figure}[]
 \centering
 \subfigure[]{
    \label{galled_tree_pex1}
    \includegraphics[scale=0.5,clip = true, viewport = -30 0 360 400]{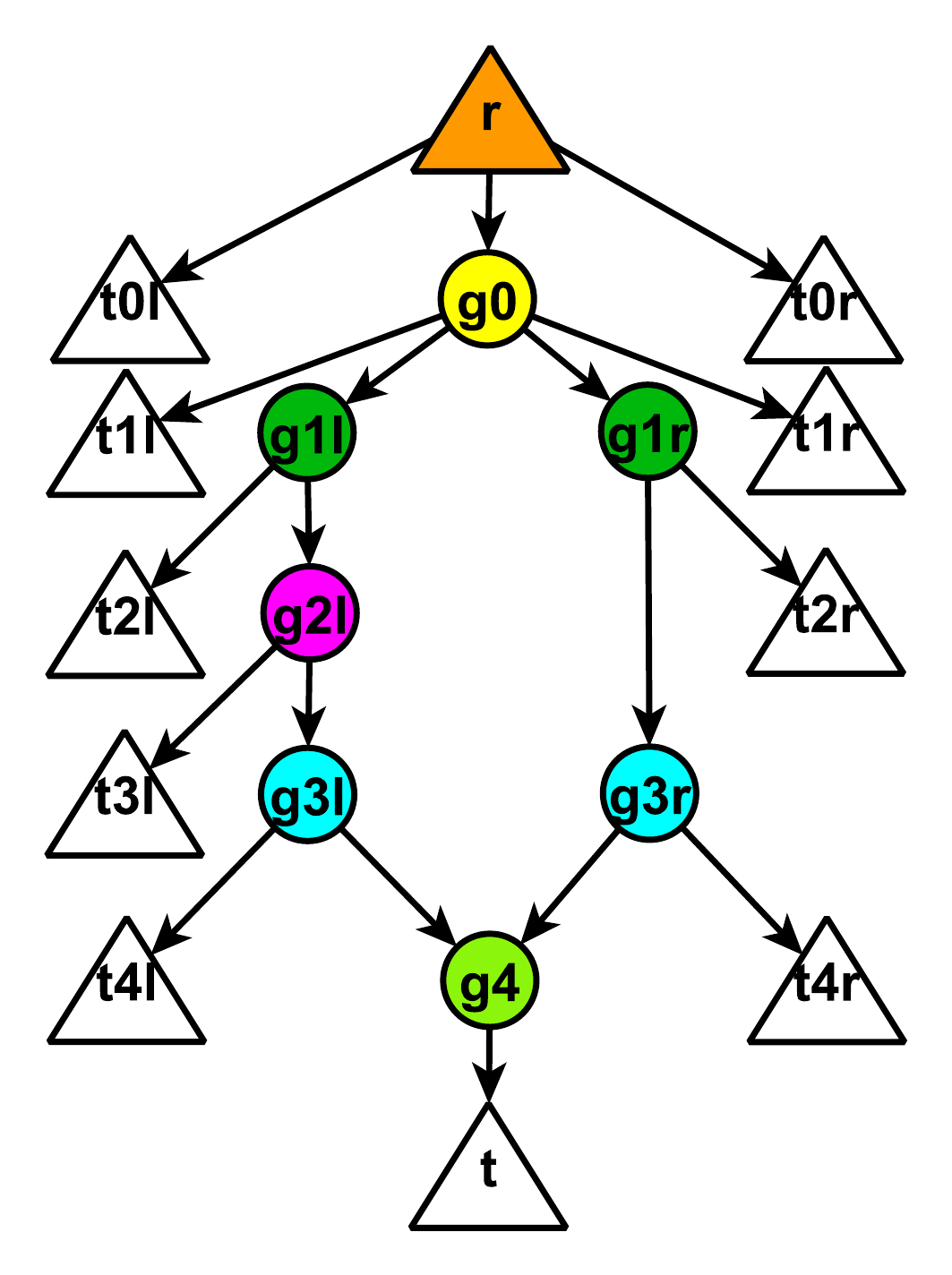}}
 \subfigure[]{
    \label{galled_tree_pex2}
    \includegraphics[scale=0.5,clip = true, viewport = 0 0 230 400]{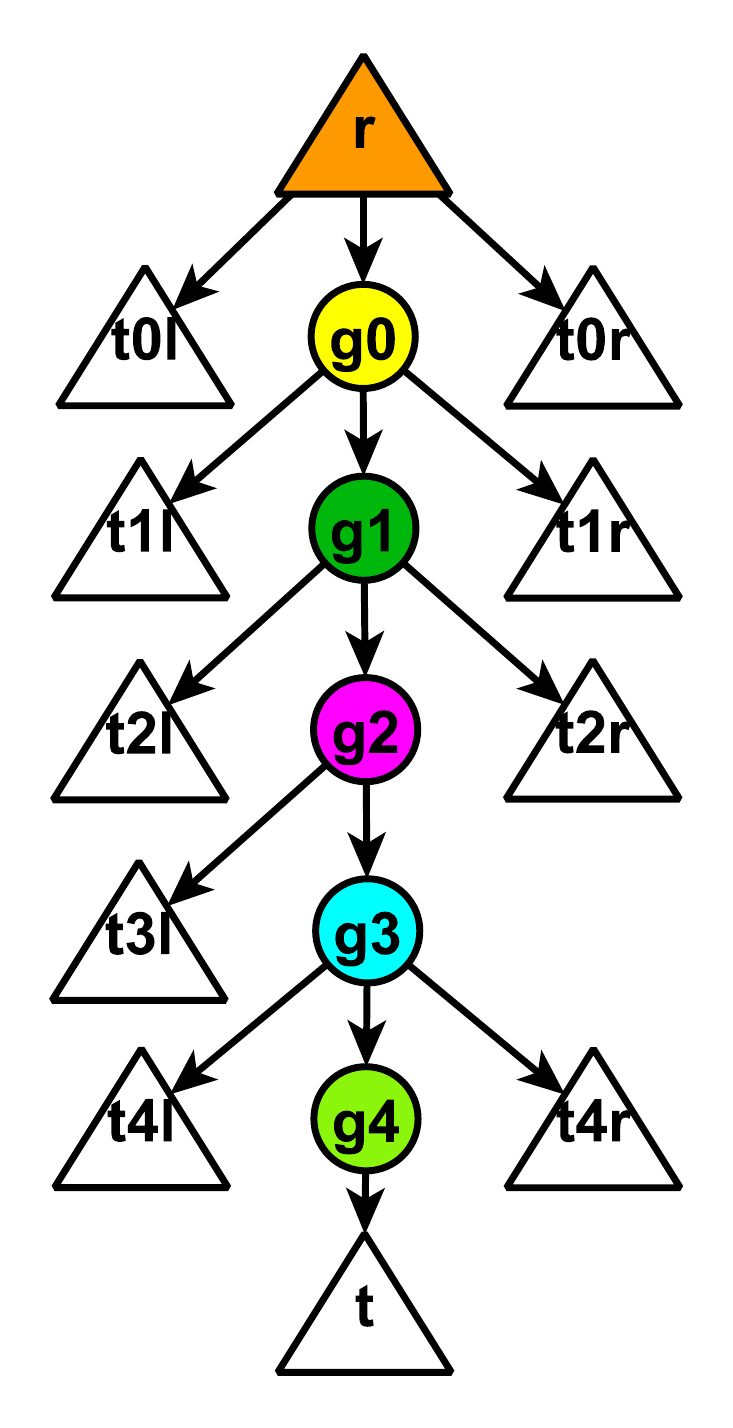}}
\caption{Transformation of a galled tree ({\bf a}) into a tree ({\bf b}). }
 \label{galled_tree_pex12}
\end{figure}

This is easy to realize considering that galled trees are almost like trees, but with some branches being made of two parallel chains, instead of one (see Figure \ref{galled_tree_pex12}).
Furthermore, this implies that the number of edges of a galled tree is $O(n)$. 

We now present an algorithm for constructing a DAGmap of a galled tree:\\
 
\newcommand{\giMath}{$g_i$}
\newcommand{\iSetMinusOne}{$i\in\{1, \ldots, k-1\}$}
\newcommand{\tiMath}{$t_i$}
\newcommand{\GMath}{$G$}
\newcommand{\TMath}{$T$}

\noindent \textbf{Algorithm 2: DAGmap drawing of galled trees}\\
\textbf{Input: } A galled tree $G$.\\
\textbf{Output: } A DAGmap drawing of $G$.

\begin{Verbatim}[commandchars=\\+\#]
1. Transform the galled tree \GMath into a tree \TMath, by unifying the two
   chains of each gall.
2. Draw the treemap of \TMath, according to the chosen treemap
   technique.
3. Split the rectangles, corresponding to the nodes of the 
   unified chains of the galls, to obtain the initial 
   parallel chains.
\end{Verbatim}

Step 1 of the above algorithm is illustrated in Figure \ref{galled_tree_pex12}. The parallel chains have been united, and nodes $g_{i_l}$, $g_{i_r}$ have been replaced by node $g_i$ while the subtrees $t_{i_l}$ and $t_{i_r}$ remain unchanged, $i\in\{1, \ldots, k\}$.

\begin{figure}[]
 \centering
 \subfigure[]{
    \label{galled_tree_pex5}
    \includegraphics[scale=0.495,clip = true, viewport = 40 550 570 850]{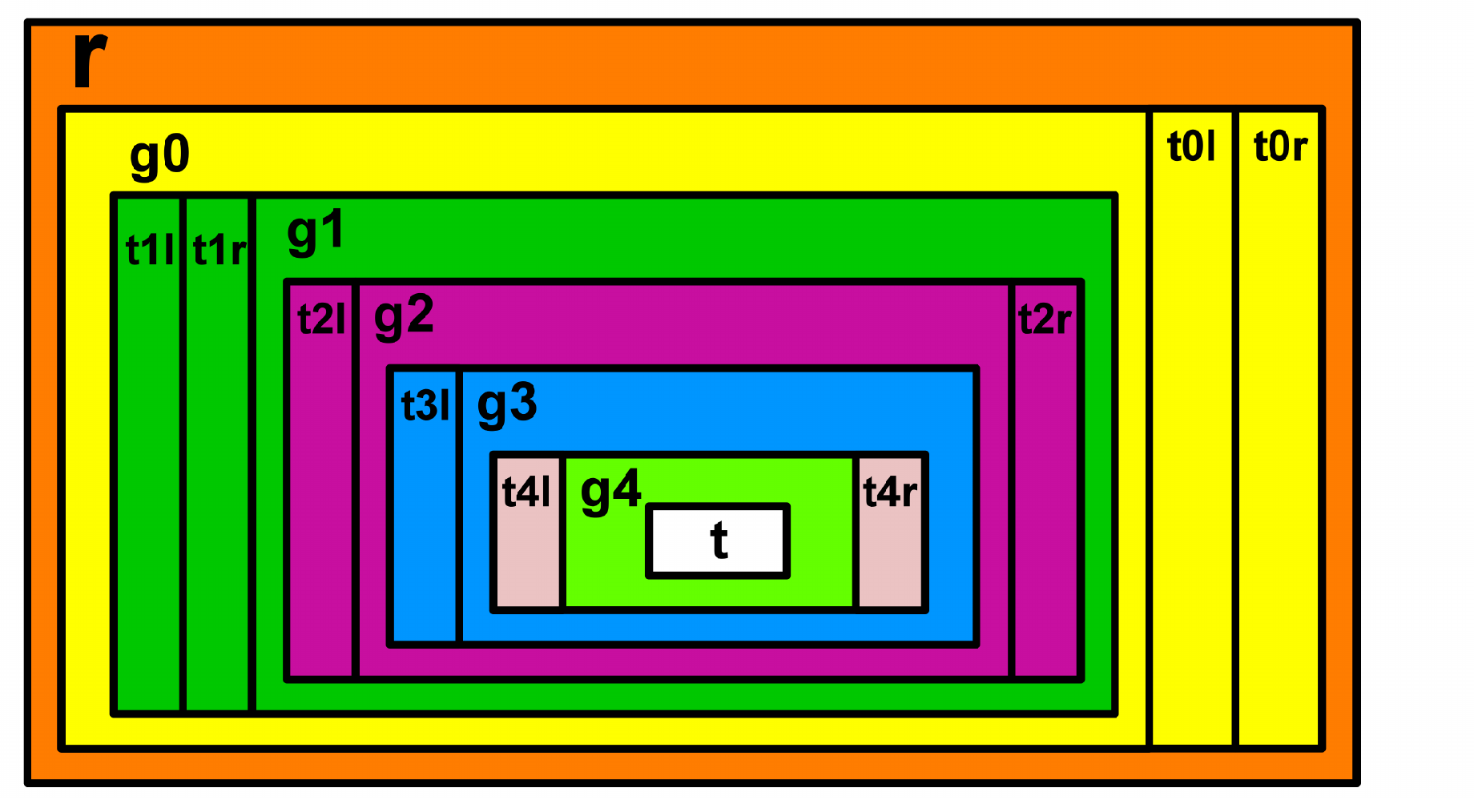}}
 \subfigure[  ]{
    \label{galled_tree_pex3}
    \includegraphics[scale=0.54,clip = true, viewport = 63 550 550 850]{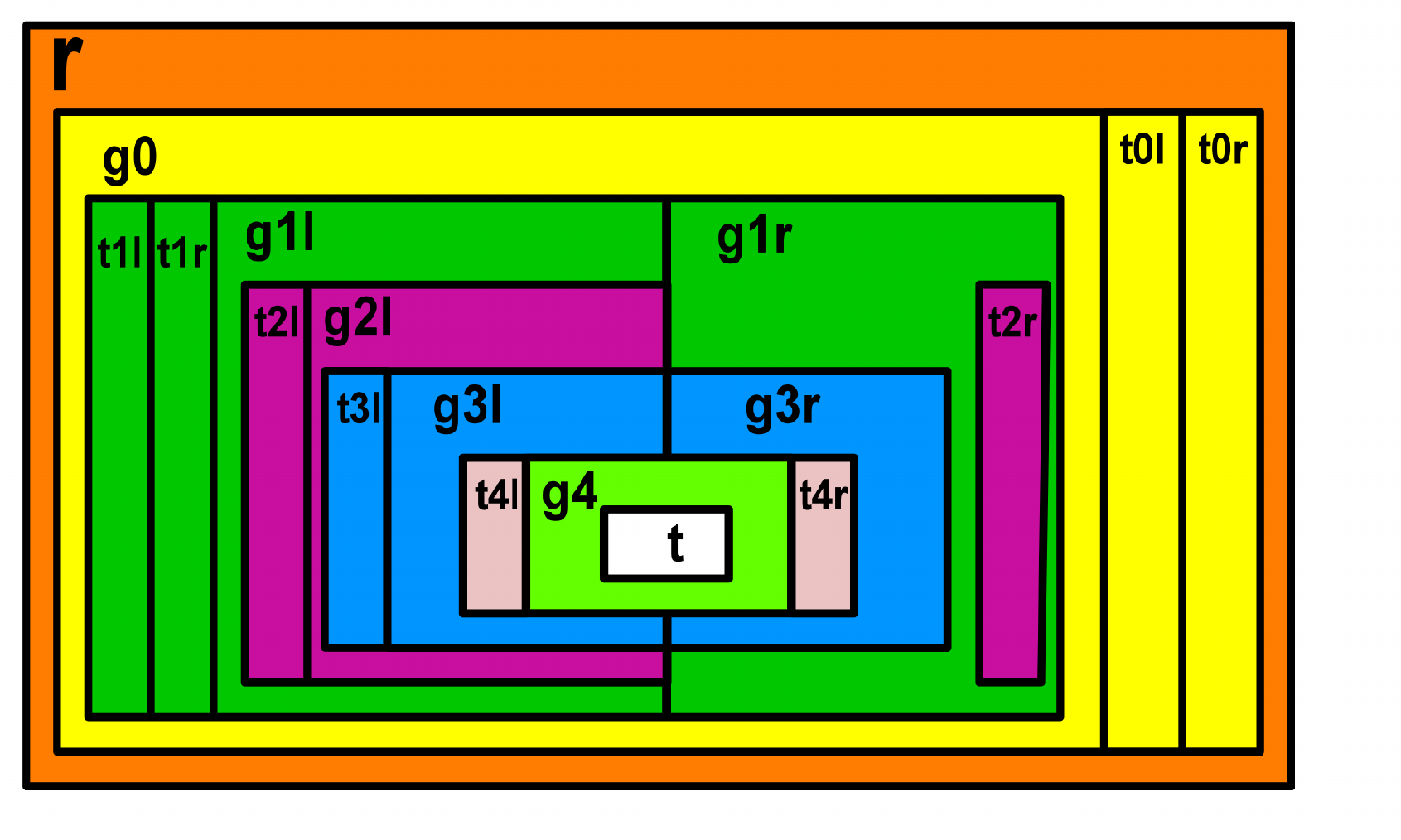}}
\caption{({\bf a}) The treemap drawing of the tree shown in Figure \ref{galled_tree_pex2}. ({\bf b}) The DAGmap drawing of the gall shown in Figure \ref{galled_tree_pex1}.}
 \label{galled_tree_pex35}
\end{figure}

The treemap of $T$, in Step 2, is drawn under the constraint that the $g_i$ nodes (which represent the union of nodes $g_{i_l}$ and $g_{i_r}$ of the DAG) must always touch both $t_{i_l}$ and $t_{i_r}$, in the same direction. This means that if we choose to place $g_{i_l}$ on the left and $g_{i_r}$ on the right, where $i\in\{1, \ldots, k-1\}$, then we will follow this convention for every $i\in\{1, \ldots, k-1\}$ (see Figure \ref{galled_tree_pex5}). Drawing the treemap of $T$ needs $O(n)$ time, if we choose a linear time layout algorithm like the slice and dice layout.

Slice and dice \cite{Johnson91} is a treemap drawing technique, 
where the initial rectangle is recursively  divided. 
The direction of each subdivision changes in each level, from horizontal to vertical.

The output of Step 3 is shown in Figure \ref{galled_tree_pex3}, where the unified nodes are split. Note that the reticulation node $g_k$ lies on both $g_{k-1_l}$ and $g_{k-1_r}$. This step needs $O(n)$ time, because in the worst case it traverses all the nodes of the graph.

From the above we conclude that:
\begin{theorem}
Every galled tree admits a DAGmap drawing, which can be computed in $O(n)$ time.
\end{theorem}

In the next section we show that galled trees can be drawn as one-dimensional DAGmaps.

\subsection{Drawing Galled Trees as One-Dimensional DAGmaps}
\label{section1DDAGmap}

A DAGmap is called one-dimensional if the initial rectangle is sliced only along the vertical
(horizontal) direction. Since the height (width) of all the rectangles is constant and equal to
the height (width) of the initial drawing rectangle, the problem is one-dimensional.

Next, we show that galled trees can be drawn as one-dimensional DAGmaps.

\begin{theorem}
	Every galled tree can be drawn as a one-dimensional DAGmap.
\end{theorem}

{\bf Sketch of Proof:} Let $G = (V, E)$ be a proper layered DAG with vertex partition $V = L_1\cup L_2 \cup \ldots \cup L_h$, where $h > 1$, such that the source (root) is in $L_h$ and the sinks are in $L_1$. 
Tsiaras \emph{ et al.} \cite{Tsiaras09} have shown that a DAG $G$ admits a one-dimensional 
DAGmap if and only if it is layered planar. 
We will show that every galled tree is layered planar, using its tree-like structure.

We transform the galled tree $G$ into a tree $T$, as shown in Figure \ref{galled_tree_pex12}. We take the vertex partition of $T$: $V_T = L_1\cup L_2 \cup \ldots \cup L_h$, where $h > 1$, such that the source (root) is in $L_h$ and the sinks are in $L_1$. Then, we define the vertex partition of the galled tree $V_G = L_1\cup L_2 \cup \ldots \cup L_h$, where $h > 1$, such that every node of $T$ which also belongs to $G$ remains at the same layer. Moreover, for every node $g_i$ of $T$ which belongs to layer $L_j$ of the partition, and is originated from the union of the nodes $g_{i_l}$ and $g_{i_r}$ of $G$, it is concluded that $g_{i_l}$ and $g_{i_r}$ will belong to layer $L_j$ of the partition $V_G$.

Since every tree is layered planar and we obtained the vertex partition of $G$ from the vertex partition of $T$, we conclude that every galled tree admits a one-dimensional DAGmap. \hspace*{\fill} \framebox \\
\\

However, not every planar galled network admits a one-dimensional DAGmap.

\begin{lemma}
	Not every planar galled network admits a one-dimensional DAGmap.
\end{lemma}

\begin{figure}
	\centering
	\includegraphics[scale=0.4,clip = true, viewport = 0 0 360 310]{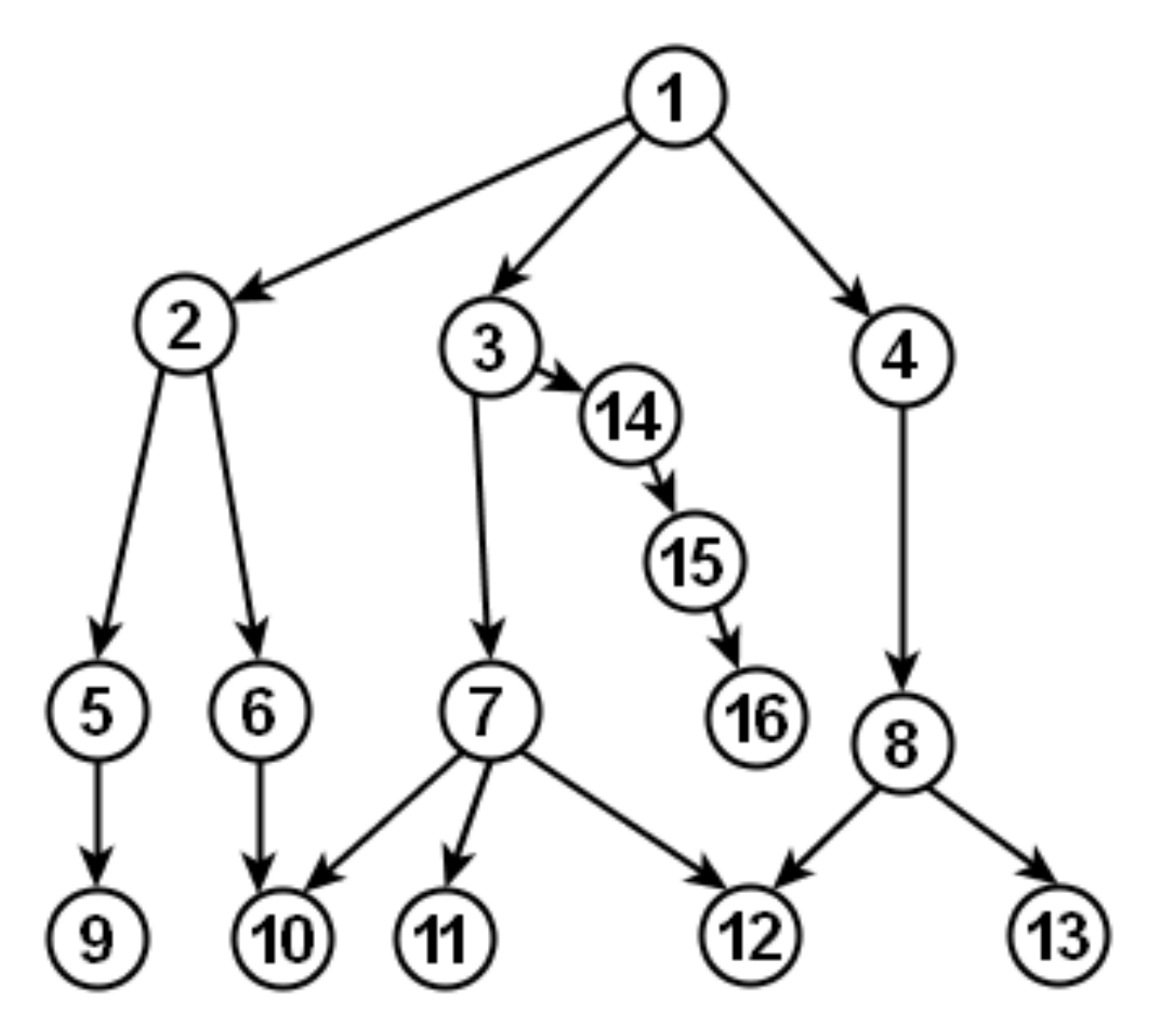}
	\caption{An example of a galled network that does not admit a one-dimensional DAGmap. }
	\label{galled_net_counterexample}
\end{figure}

{\bf Sketch of Proof:} In Figure \ref{galled_net_counterexample} an example of such a planar galled network is shown, that does not admit a one-dimensional DAGmap. Node $16$ will not be able to be drawn in the line of level $4$ without edge crossings. However, as it will be shown in the next section, this graph can be DAGmap drawn. \hspace*{\fill} \framebox 

\section{DAGmaps for Galled Networks}
\label{dagmaps4GN}

In this section we investigate how to draw galled networks as DAGmaps.
From Theorem \ref{NP} we have that this problem is NP-complete. 
Therefore, it is worth examining the problem of drawing planar galled networks as DAGmaps.
In the following lemma we show that planar galled networks are a subset of the set of galled networks.    

\begin{figure}[]
\begin{center}
 \subfigure[ ]{
    \label{galled_network}
    \includegraphics[scale=0.5,clip = true, viewport = 0 0 330 220]{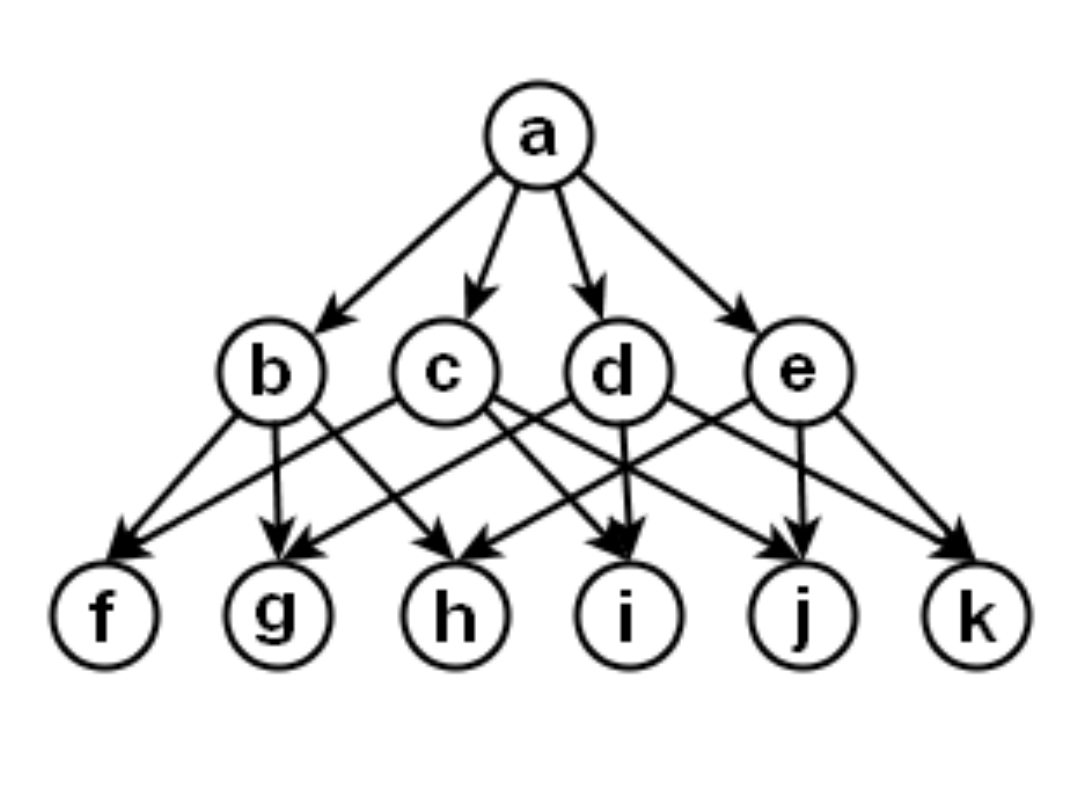}}
 \subfigure[]{
    \label{galled_network_k5}
    \includegraphics[scale=0.5,clip = true, viewport = 10 0 300 220]{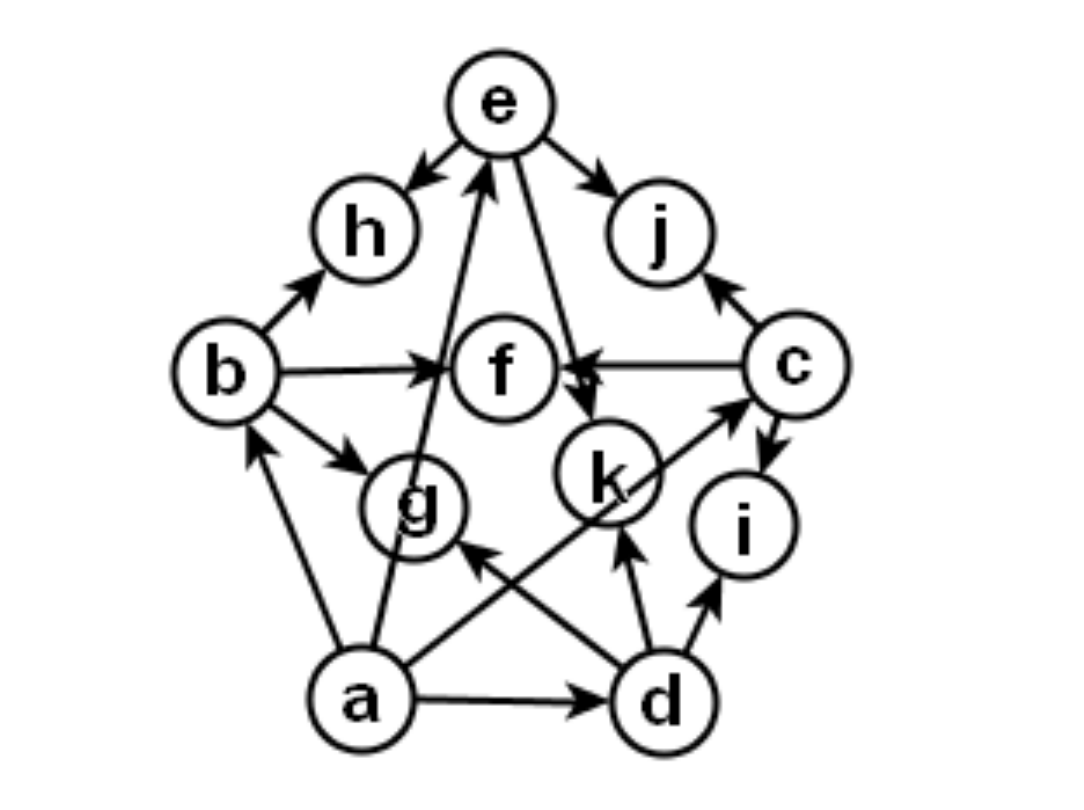}}
\caption{An example of a non planar galled network. As we can see the network ({\bf a}) 
	is the same with the network ({\bf b}), which is topologically equivalent to $K_5$. }
 \label{galled_network_not_planar}
\end{center}
\end{figure}

\begin{lemma}
Not every galled network is planar.
\end{lemma}
\noindent
{\bf Sketch of Proof:} 
This lemma can be proved by creating a family of galled networks that contain a subgraph homeomorphic to $K_5$ \cite{Kuratowski30}. Figure $\ref{galled_network_not_planar} (a)$ depicts a Galled network. 
This is a non planar galled network since it is topologically the same as the network shown in Figure $\ref{galled_network_not_planar} (b)$, which is homeomorphic to $K_5$. \hspace*{\fill} \framebox \\
\\

Since planar galled networks represent phylogenetic networks, it is clear that all edges flow 
in the same direction monotonically. This means that planar galled networks are upward (downward) planar graphs. 
Therefore, we have the following:

\begin{fact}
	Planar galled networks are upward planar.
\end{fact}

By definition, the phylogenetic networks are single source directed acyclic graphs.
Therefore, we have the following:
  
\begin{fact}
	Each planar galled network is a single source upward planar directed acyclic graph.
	\label{fact4}
\end{fact}

In order to draw planar galled networks as DAGmaps, without node duplication, 
we will relax the rule for drawing DAGmaps, which states that every node is drawn as a rectangle. 
Specifically, we will allow nodes to be drawn as rectilinear cohesive polygons.
Next, we present an algorithm that produces DAGmaps of planar galled networks. \\

\noindent \textbf{Algorithm 3: DAGmap drawing of planar galled networks}\\
\textbf{Input: } A planar galled network $G$.\\
\textbf{Output: } A DAGmap drawing of $G$.

\newcommand{\GTMath}{$GT$}
\newcommand{\rMath}{$r$}

\begin{Verbatim}[commandchars=\\+\#]
1. Transform the galled network \GMath into a galled tree \GTMath, by
   splitting the nodes that belong to more than one galls, so as
   no gall shares its nodes with other galls.
2. Order all subtrees of \GTMath such that:
3.   The nodes created by the splitting of nodes of \GMath are moved 
     so that they are adjacent to each other.
4. Draw the DAGmaps of the galls of \GTMath. Nested  galls are drawn 
   recursively.
5.  Unify the split nodes and remove unused space.
\end{Verbatim}

\begin{figure}[]
	\centering
	\subfigure[]{
		\label{alg3fig1}
		\includegraphics[scale=0.35,clip = true, viewport = 220 160 1000 480]{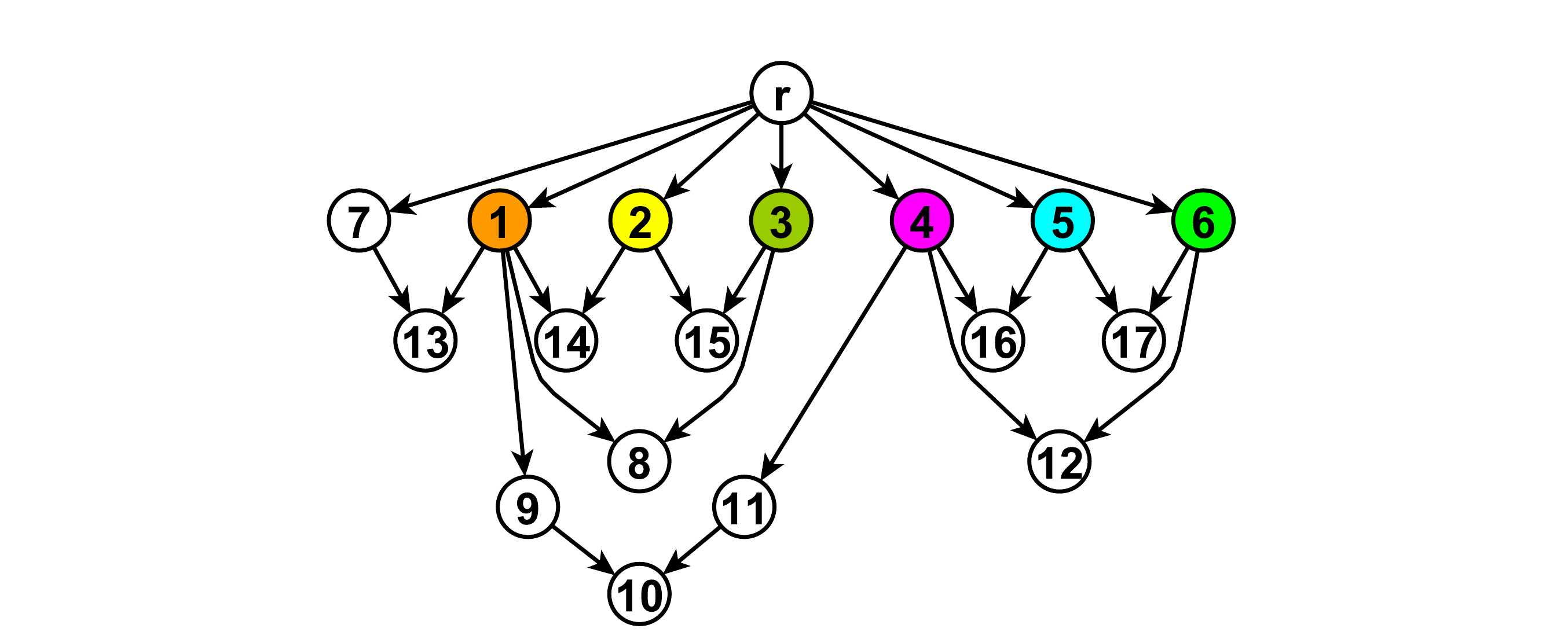}}
	\subfigure[]{
		\label{alg3fig2}
		\includegraphics[scale=0.35,clip = true, viewport = 125 160 1100 480]{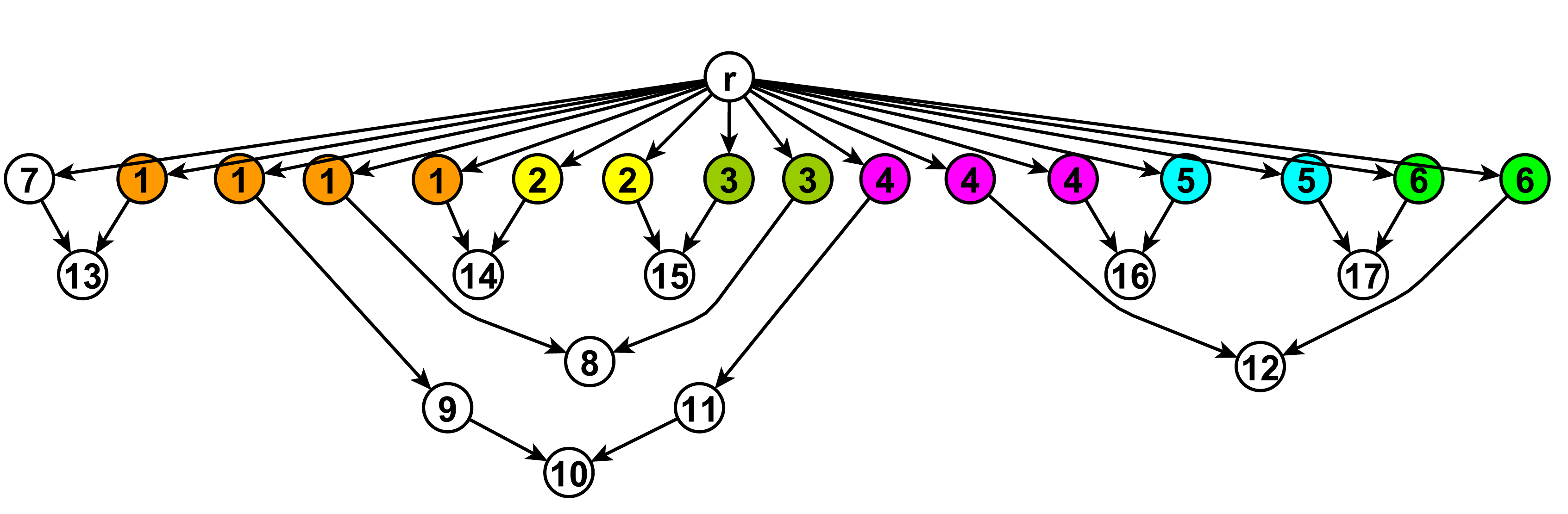}}
	\caption{Transformation of a galled network ({\bf a}) to a galled tree ({\bf b}).}
	\label{alg3fig12}
\end{figure}

Step 1 of the above algorithm is illustrated in Figure \ref{alg3fig12}. As shown, every node $u$ that participates in $k$ galls ($k > 1$) is being replaced by $k$ nodes $u_i$, $i\in\{1, \ldots, k\}$. Each node $u_i$ participates in only one gall. Consequently $GT$ is a galled tree because there is no gall that shares nodes with any other gall. This step needs $O(n)$ time, because in the worst case it traverses all the nodes of the graph, and the number of edges of a planar graph is $O(n)$.

In Steps 2 and 3 we define the order of all subtrees of the galled tree \GTMath. 
The goal is to find an ordering such that all splitted nodes are neighbors. 
We observe that a proper nesting of the galls produces a planar embedding of \GMath.
Thus, given a planar embedding $\Gamma$ of \GMath, it is easy to find the correct order of all subtrees.
Specifically,  the order of the subtrees of {\GTMath}  is determined by the 
clockwise order of the incoming and outgoing edges of each node (to be splitted) in $\Gamma$.
Bertolazzi \emph{et al.} \cite{79626} have shown that if a single source digraph is upward planar, 
then its drawing can be constructed in $O(n)$ time. 
Thus, given Fact \ref{fact4}, we can produce an upward planar drawing of a planar galled network in linear time.

\begin{figure} []{
		\centering
		\includegraphics[scale = 0.4,clip = true, viewport = -50 30 700 560]{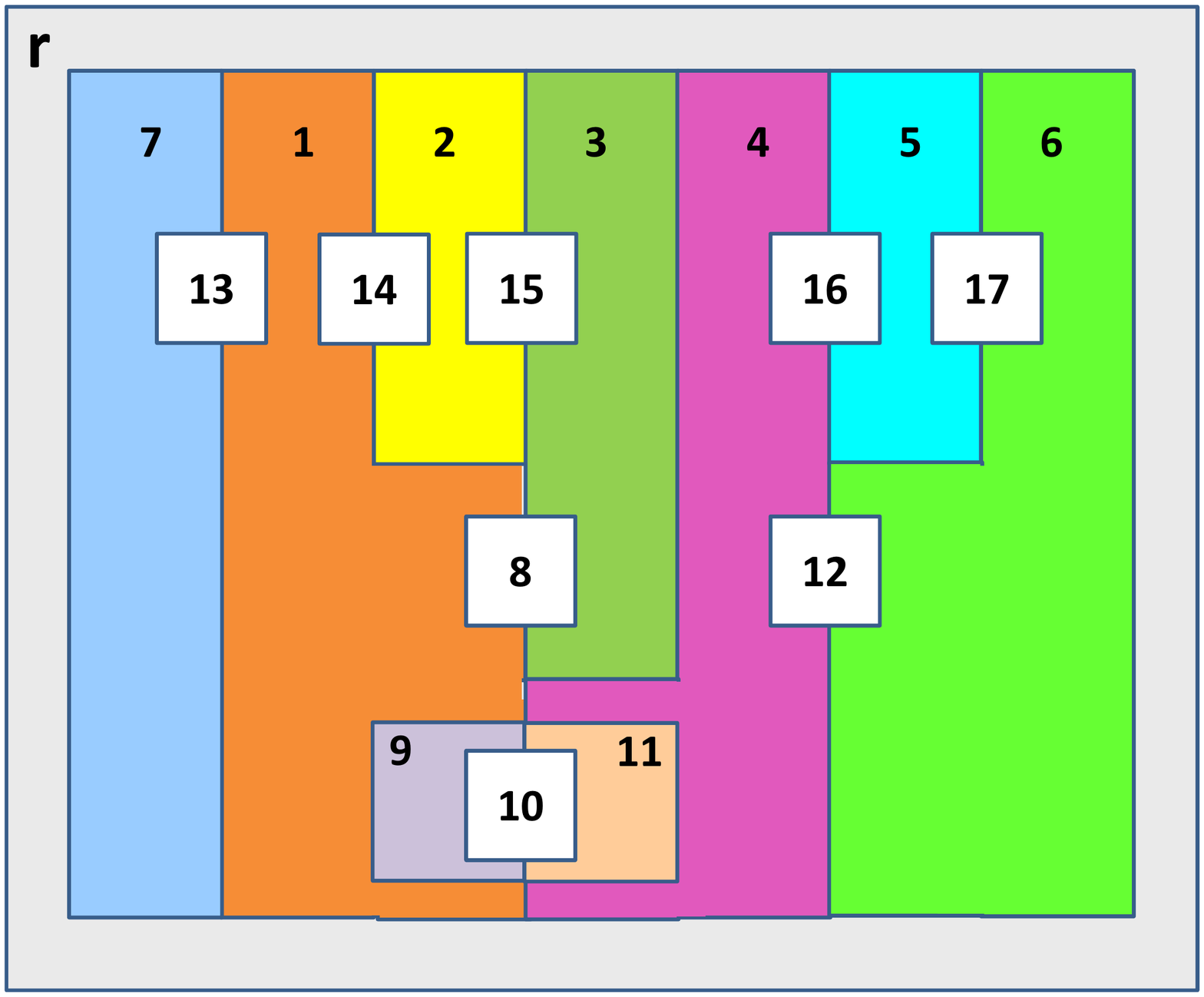}}
	\caption{ The DAGmap drawing of the galled network of Figure \ref{alg3fig1} produced by Algorithm 3.}
	\label{alg3}
\end{figure}

The drawings of the DAGmaps of the galls of $GT$ (Step 4) are obtained by executing Algorithm 2. 
The running time of this algorithm is $O(n)$. 
Finally, the unification of Step 5 needs $O(n)$ time in the worst case, since it is the reverse procedure of Step 1. The output is shown in Figure \ref{alg3}.


Generally speaking, the node splitting process triggers the duplication of all of its out-neighbors. 
Therefore, the transformation of a DAG into a tree leads to trees with 
(potentially exponentially) many more nodes than the original DAG.
However, the node splitting of Step 1 does not have the exponential effects of the 
ordinary node duplication, since all the duplicated nodes of this case are neighbors.
 From the above, we realise that Algorithm 4 takes $O(n)$ time, and combining this with Algorithm 3, we conclude that:

\begin{theorem}
Every planar galled network admits a DAGmap drawing, which can be computed in $O(n)$ time.
\end{theorem}

\section{Conclusions and Future Work}
DAGmaps, an extension of Treemaps, represent an effective space filling 
visualization method to display and analyze hierarchical data. 
In this paper we have presented algorithms that use DAGmap drawings for 
the visualization of two categories of phylogenetic networks, galled trees and planar galled networks.
Future work will cover the study of more categories of phylogenetic networks, 
in addition to answering the question whether one could minimize the number of node duplications 
performed during Step 1 of Algorithm 3 in the case of non planar galled networks.
Furthermore, we intend to develop a visualization tool for processing phylogenetic networks
and displaying them as DAGmaps.



\section*{Acknowledgments}
We thank Irini Koutaki-Pantermaki who contributed some ideas in an earlier version of the paper, and Vassilis Tsiaras for useful discussions.
%
%

\bibliographystyle{plain}
\bibliography{biblio}

\end{document}